\documentclass{aastex}

\newcommand{\kmps}{km~s$^{-1}$} 
\newcommand{\phoe}{{\tt PHOENIX}}
\newcommand{\snia}{SN~I\lowercase{a}}
\newcommand{\sneia}{SNe~I\lowercase{a}} 
\newcommand{\cdd}{CS15DD3}
\newcommand{\dd}{DD21\lowercase{c}}

\bibstyle{apj} 
\begin{document}

\title{SN~1984A and Delayed Detonation Models of Type~I\lowercase{a}
Supernovae}

\author{Eric~J. Lentz, E.~Baron, David Branch} \affil{Department of
Physics and Astronomy, University of Oklahoma, 440 W. Brooks St.,
Norman, OK 73019-0225} \email{lentz,baron,branch@mail.nhn.ou.edu}
\author{and} \author{Peter~H. Hauschildt} \affil{Department of Physics
and Astronomy \& Center for Simulational Physics, University of Georgia,
Athens, GA 30602-2451} \email{yeti@hal.physast.uga.edu}

\begin{abstract}

SN~1984A shows unusually large expansion velocities in lines from
freshly synthesized material, relative to typical Type~Ia Supernovae
(\sneia).  SN~1984A is an example of a group of \sneia\ which have
very large blue-shifts of the P-Cygni features, but otherwise normal
spectra.  We have modeled several early spectra of SN~1984A with the
multi-purpose NLTE model atmosphere and spectrum synthesis code,
\phoe.  We have used as input two delayed detonation models: \cdd\
\citep{iwamoto99} and \dd\ \citep{hwt98}. These models show line
expansion velocities which are larger than that for a typical
deflagration model like W7, which we have previously shown to fit the
spectra of normal \sneia\ quite well. We find these delayed detonation
models to be reasonable approximations to large absorption feature
blue-shift \sneia, like SN~1984A. Higher densities of newly
synthesized intermediate mass elements at higher velocities, $v >
15000$~\kmps, are found in delayed detonation models than in
deflagration models.  We find that this increase in density at high
velocities is responsible for the larger blue-shifts in the synthetic
spectra.  We show that the variations in line width in observed
\sneia\ are likely due to density variations in the outer,
high-velocity layers of their atmospheres.

\end{abstract}

\clearpage

\section{Introduction}

The current understanding of the progenitor system for \sneia\ features
a sub-Chandrasekhar mass C+O  white dwarf accreting mass from a
companion star until reaching the Chandrasekhar mass, or by merging with
another C+O white dwarf and then exploding \citep[for reviews see][and
references therein]{livio00,prog95}. Most \sneia\ show a correlation
between luminosity and light curve shape, but some \sneia\ show
uncorrelated spectral properties \citep{bvdb93,hathighv00}. Events
like SN~1984A \citep{bran84a87,barbon84a89} and SN~1997bq
\citep{garn97bq00} show unusually large blue-shifts in the spectra
\citep[for more see][]{hathighv00}.  These \sneia\ have otherwise
normal properties and  the usual spectral features.  They only appear
different under close inspection.  We have modeled SN~1984A using
delayed detonation models, in which an initially sub-sonic flame
(deflagration) transitions to a super-sonic shock (detonation).  We have
shown that the parameterized deflagration model, W7 \citep{nomw7}, fits
normal \sneia\ quite well \citep{l94d00}. However, we found that
delayed detonation models have abnormally large blue-shift velocities in
their synthetic spectra.  Thus, these delayed detonation models would
seem to be ideal for \sneia\ with abnormally high expansion velocities.

\section{Computational Procedures}

\subsection{\phoe}

We have used the multi-purpose spectrum synthesis and model atmosphere
code \phoe~{\tt 10.8}\ \citep[see][and references therein]{hbjcam99}.
\phoe\ has been designed to accurately include the various effects of
special relativity important in rapidly expanding atmospheres, like
supernovae. Ionization by non-thermal electrons from gamma-rays from the
nuclear decay of $^{56}$Ni that powers the light curves of \sneia\ is
taken into account.  We have updated the method for calculating the
gamma-ray deposition from the single $\Lambda$-iteration technique of
\citet{nugphd} \citep[based on the method of][]{sw84}, to a solution of
the spherically symmetric radiative transfer equation for gamma-rays
with \phoe\ \citep{lentzphd}. We used an effective gamma-ray opacity,
$\kappa_{\gamma} = 
0.06$~Y$_e$~cm$^2$~g$^{-1}$ \citep{cpk80}. 
In the models presented
here we solve the NLTE rate equations for \ion{H}{1}~(10 levels/37
transitions), \ion{He}{1--ii}~(19/37, 10/45),
\ion{C}{1}~(228/1387), \ion{O}{1}~(36/66), \ion{Ne}{1}~(26/37),
\ion{Na}{1}~(53/142), \ion{Mg}{2}~(73/340),
\ion{Si}{2}~(93/436), \ion{S}{2}~(84/444), \ion{Ti}{2}~(204/2399),
\ion{Fe}{2}~(617/13675), and \ion{Co}{2}~(255/2725).

\subsection{\cdd}

The model \cdd\  was prepared from the hydrodynamical calculation of
\citet{iwamoto99} by homologous expansion of the hydro model and by
assuming a rise time of 20 days after explosion to maximum light
\citep[e.g.,][]{reissetalIaev00,akn00}.  A few layers near $\sim 6000$
\kmps\ having non-monotonic velocities have been re-mapped, while we
have included the corresponding density spike from the hydrodynamic
interactions.  The models, which extend to velocities $>
100,000$~\kmps, have been truncated at 40,000~\kmps\ for the
pre-maximum spectra (13, 15, and 17 days after explosion) and for the
post-maximum spectrum (28 days after explosion) to save model points
to resolve parts of the atmosphere which are relevant to spectral
formation.  In tests, we have found that the excluded portions of the
atmosphere do not affect the spectra. At each epoch, we have fit the
luminosity to match the shape and color of the observations, while
solving for the energy balance and converged NLTE rate equations.

\subsection{\dd}

The model \dd\ \citep{hwt98} has also been homologously expanded to
the epoch of the observations with the same 20 day rise time.  The
hydrodynamical models have been extended from the largest velocity of
$\sim 25,000$~\kmps\ to 30,000~\kmps\ using a steep power law and the
same composition as the last model layer.  Because of the steepness of
the power law the extra matter did not affect the spectra in
tests. The extension was necessary to provide a better outer structure
for the atmosphere to increase numerical stability.  We have used the
luminosities provided in \citet{hwt98} for \dd\ without modification
or fitting, thus directly including the effects of time-dependence in
the SN atmosphere.

\section{Results}

\begin{figure} 
\begin{center} 
\leavevmode
\epsscale{0.95}
\plotone{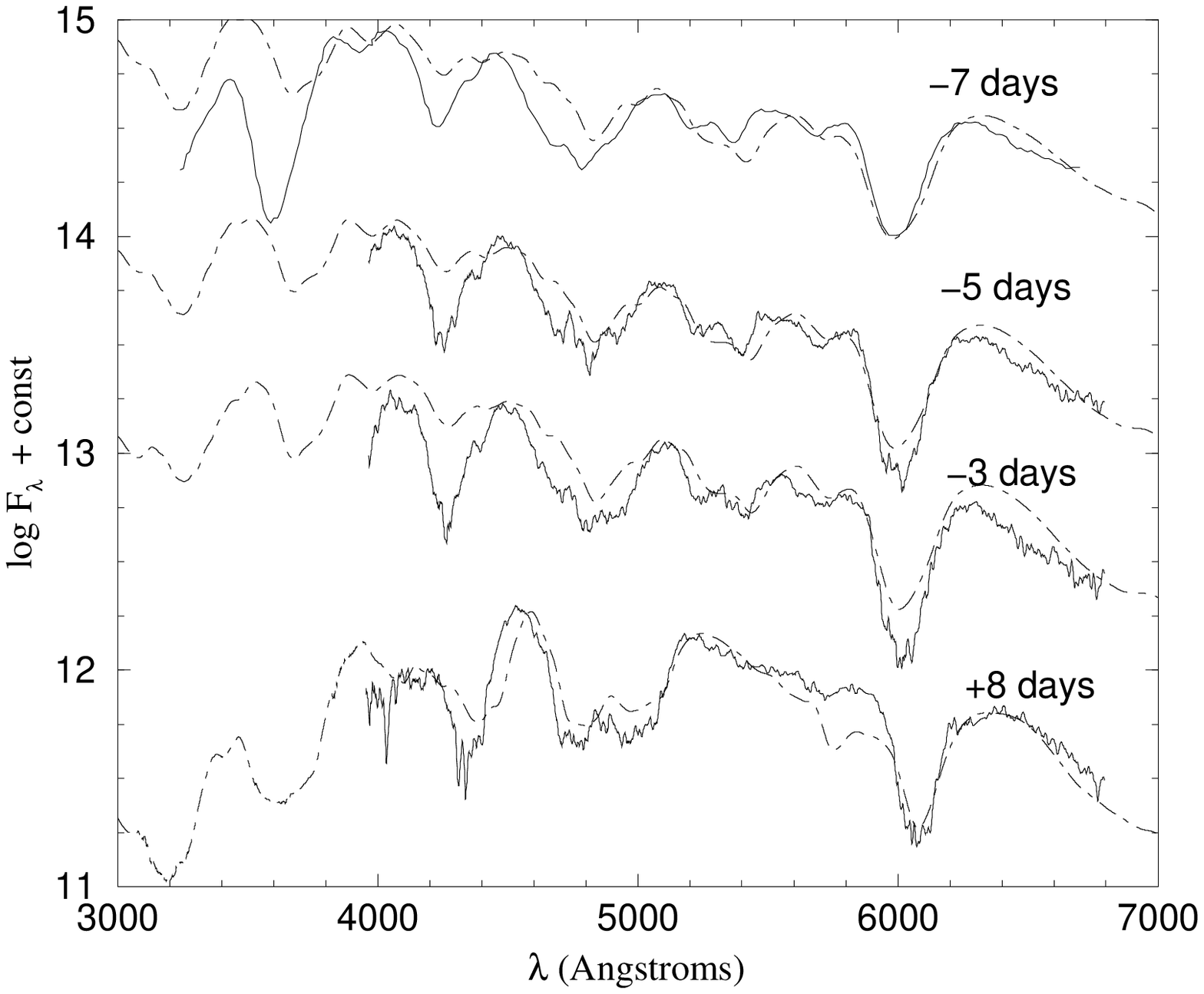}
\end{center}
\caption{Synthetic spectra
for \dd~(dot-dashed lines) plotted against observed spectra for
SN~1984A~(solid lines).\label{fig:dd21c}} 
\end{figure}

We have plotted the synthetic spectra from \dd\ for days 13, 15, 17, and 28 
after explosion against the observations from -7, -5, -3, and +8 days 
relative to maximum light respectively in Figure~\ref{fig:dd21c}. 
We use the -7 day spectrum from \citet{wegmcm87}.  The 
epoch of maximum light and remaining spectra are from \citet{barbon84a89}. The
spectra fit remarkably well. The excellent agreement of the synthetic
spectral color to the observations shows both that the calculated bolometric
luminosities in \citet{hwt98} are accurate, and  that
\phoe\ does an excellent 
job of reproducing time-dependent calculations using the externally
calculated bolometric luminosities. The \ion{Si}{2} feature near 6000~\AA, the
defining characteristic of \sneia, fits very well in both velocity
(wavelength) range and shape. The general fit of the 28 day synthetic
spectrum over 4500--5600~\AA\ is quite good, especially since the model is
making the transition between optically thick and optically thin and
therefore is more difficult to model. The \ion{S}{2}
feature at 5100--5400~\AA\ is shifted a little to the red in the
pre-maximum spectra, indicating that the sulfur in the model should be
moving faster to fit SN~1984A. The 4400~\AA\ and 5000~\AA\ \ion{Fe}{2}
features in the pre-maximum spectra are both too weak, but seem to have
the correct velocity. The \ion{Ca}{2}~H+K feature in the day 13 spectrum
is both a bit too weak and too slow.  This indicates that higher
abundance of calcium may be needed at these extreme velocities. The
\ion{Ca}{2}~H+K feature is also quite sensitive to the model
temperature.  The velocity extent of the model is large enough to form
the observed \ion{Ca}{2}~H+K line blue-shift.

\begin{figure} 
\begin{center} 
\leavevmode
\epsscale{0.95}
\plotone{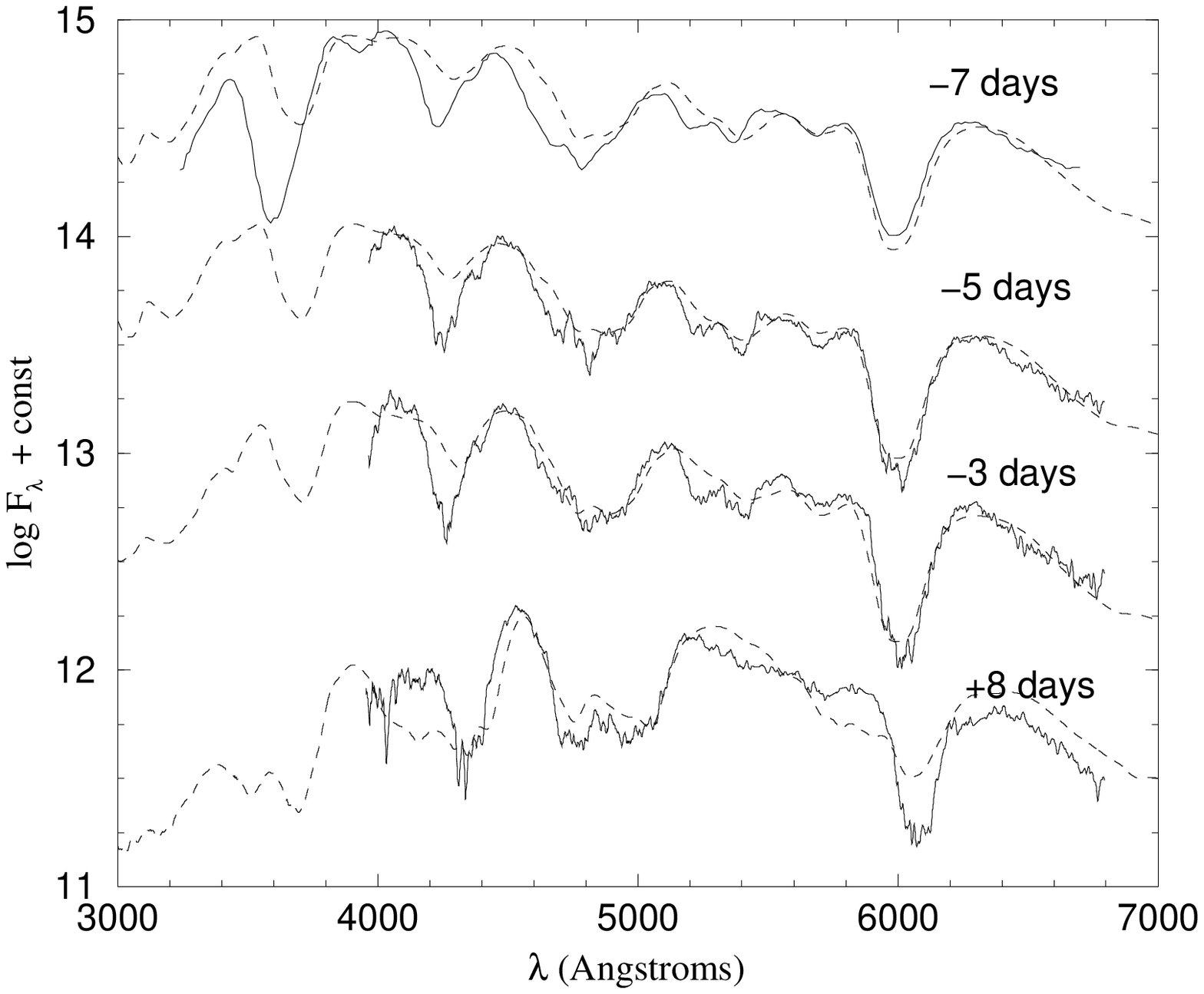}
\end{center}
\caption{Synthetic spectra
for \cdd~(dashed lines) plotted against observed spectra for
SN~1984A~(solid lines).\label{fig:cdd3}} 
\end{figure}

Our best fits, adjusting the total bolometric luminosity to fit the color, to
SN~1984A with \cdd\ are plotted in Figure~\ref{fig:cdd3}. Again the fit
to the \ion{Si}{2} feature at 6000~\AA\ is excellent. The \ion{S}{2} in
the pre-maximum spectra shows improvement over \dd, but is a little slow
on day 13 and weak on day 17. The 5000~\AA\ \ion{Fe}{2} feature fits
much better in all synthetic spectra than \dd, especially in showing the
broad, deep-bottomed shape of the feature. The pre-maximum 4400~\AA\
feature again is a bit weak and possibly a bit slow on day 13.  The
\ion{Ca}{2}~H+K feature shows the same weak and slow line as \dd. The
day 28 spectrum fits fairly well except for the depth of the 6100~\AA\
\ion{Si}{2} feature and the small continuum deficit just blueward of
that feature. This model has a continuum optical depth, $\tau_{5000} =
1.7$, which is quite low. 

\begin{figure} 
\begin{center} 
\leavevmode
\epsscale{0.95}
\plotone{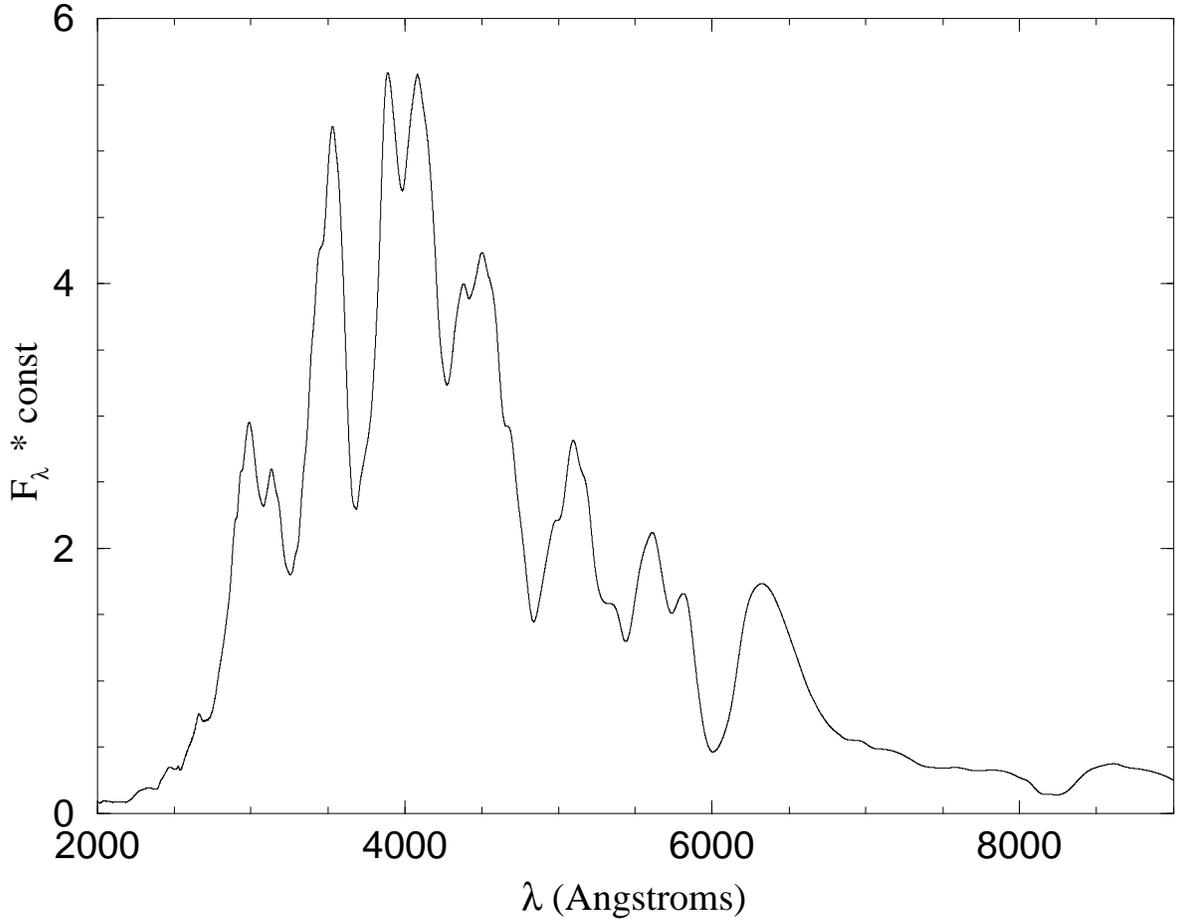}
\end{center}
\caption{Synthetic spectra
for \cdd~(dashed lines) plotted on the same scale of Figure~8 of HWT98.
\label{hwtcompare}} 
\end{figure}

Comparing our results to those of Figure~8 in \citet{hwt98} (HWT98), the
general features are the same, but the velocities of the lineshapes in
HWT98 appear significantly slower than in our synthetic
spectra. Figure~\ref{hwtcompare} displays the 17~day spectrum in a
format that facilitates direct comparison with Figure~8 of
HWT98. There are differences in the Ca~II IR triplet, which is 
significantly stronger in HWT98, also the silicon ratio
\citep{nugseq95} is much larger in our calculations. We also find
significantly more flux near 4000~\AA\ than do HWT98.

\begin{figure} 
\begin{center} 
\leavevmode
\epsscale{0.95}
\plotone{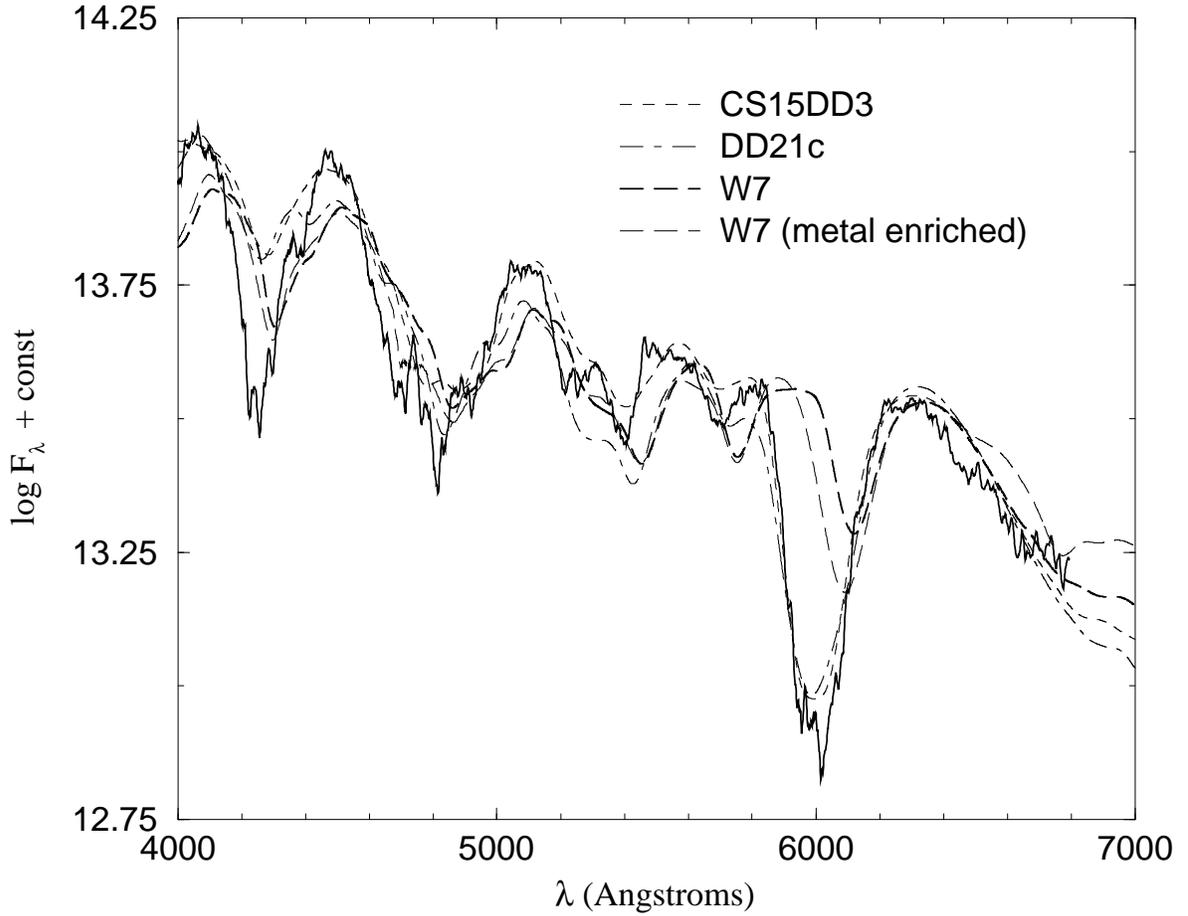}
\end{center}
\caption{Day 15 synthetic
spectra for \cdd~(short-dashed line), W7~(thick long-dashed line), W7 with
the metallicity in the unburned C+O layer enhanced by a factor of 10
\protect\citep{lentzmet00} (thin long-dashed line),
and
\dd~(dot-dashed line) plotted against observed -5~day spectrum for
SN~1984A~(solid lines). Clearly, metallicity effects cannot produce
the extremely high velocities seen here.\label{fig:day15}} 
\end{figure}

\begin{figure} 
\begin{center} 
\leavevmode
\epsscale{0.95}
\plotone{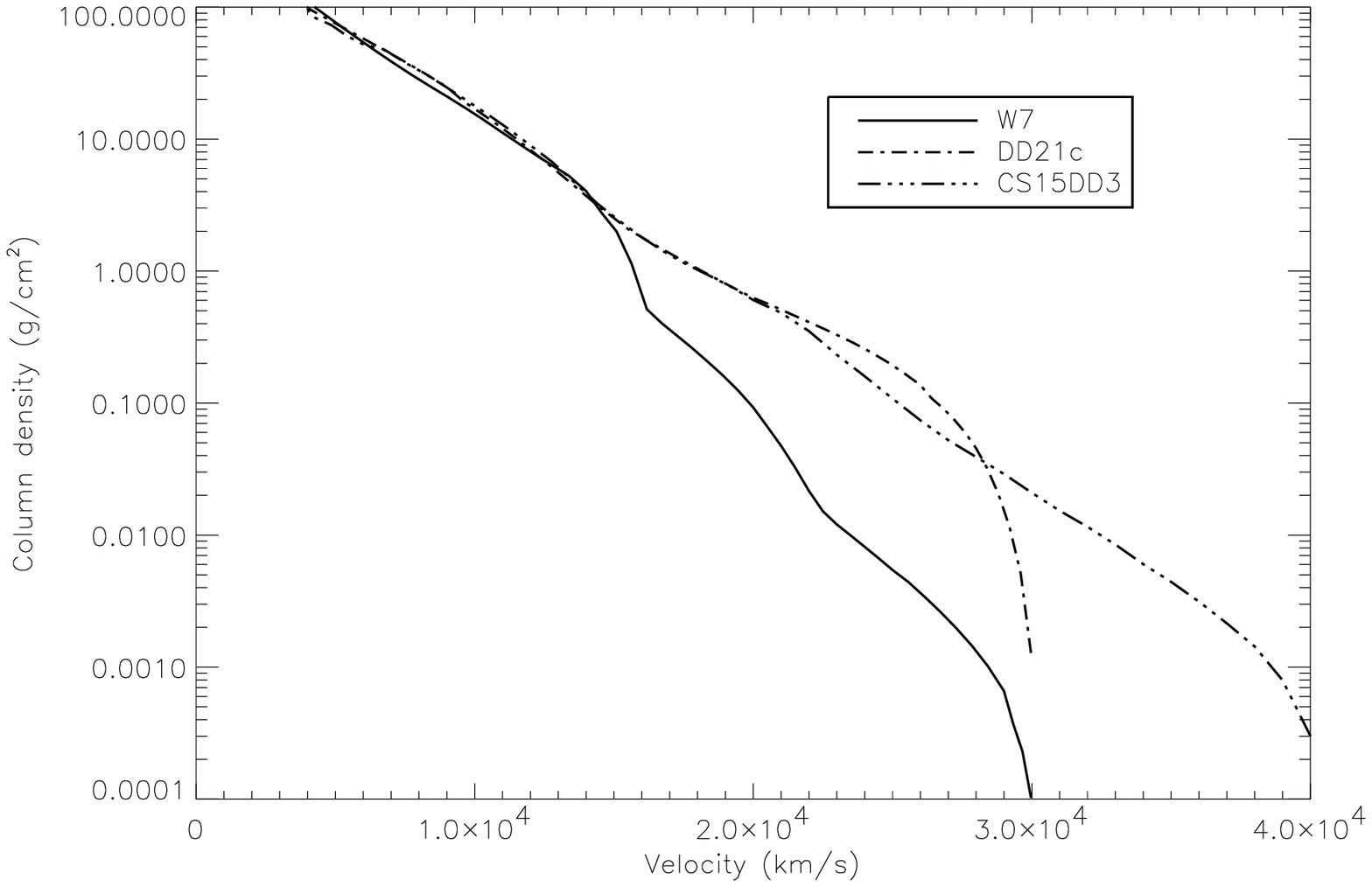}
\end{center}
\caption{Column density for
\cdd~(triple dot-dashed line), W7~(solid line), and \dd~(dot-dashed
line) plotted against velocity.\label{fig:colden}} 
\end{figure}

To demonstrate the similarities between the two delayed detonation
models and the contrasts to the deflagration model, W7, we have
plotted the day 15 spectra for \dd, \cdd, and W7 with the -5~day
SN~1984A spectrum in Figure~\ref{fig:day15}.  We have also plotted W7
where we have artificially enhanced the metallicity of the unburned
C+O layer by a factor of 10 \citep[see][where we discussed how
progenitor metallicity variation could affect the silicon
line]{lentzmet00}, clearly even an extreme value for the metallicity
can not be the cause of the high velocities of SN~1984A.  We can see
that although \cdd\ is a bit better in the \ion{Fe}{2}, \ion{S}{2},
and 5700~\AA\ \ion{Si}{2} features, the two delayed detonation models are otherwise quite
similar. The synthetic W7 spectrum, which has been carefully fit to
the ordinary \snia\ SN~1994D \citep{l94d00}, has the right color, but
the spectral features are all too slow. To understand these
differences, we have plotted the integrated column density for all
three models in Figure~\ref{fig:colden}.  The delayed detonation
models show larger column density for $v > 15,000$~\kmps, but are like
W7 at slower velocities.  In W7, the high velocity region is
unburned C+O, but in the delayed detonation models it also includes
some freshly synthesized intermediate mass elements including calcium,
sulfur, and silicon.  It is this increased density of intermediate
mass elements at high velocities to which we attribute  the high
blue-shift features in \sneia\ like SN~1984A.  We suggest that the
variations in $v_{10}$(\ion{Si}{2}), the blue-shift velocity of the
main \ion{Si}{2} feature 10 days after maximum light, noted in
\citet{hathighv00} may be due to higher densities in the outermost
layers of the supernova.

\begin{figure} 
\begin{center} 
\leavevmode
\epsscale{0.95}
\plotone{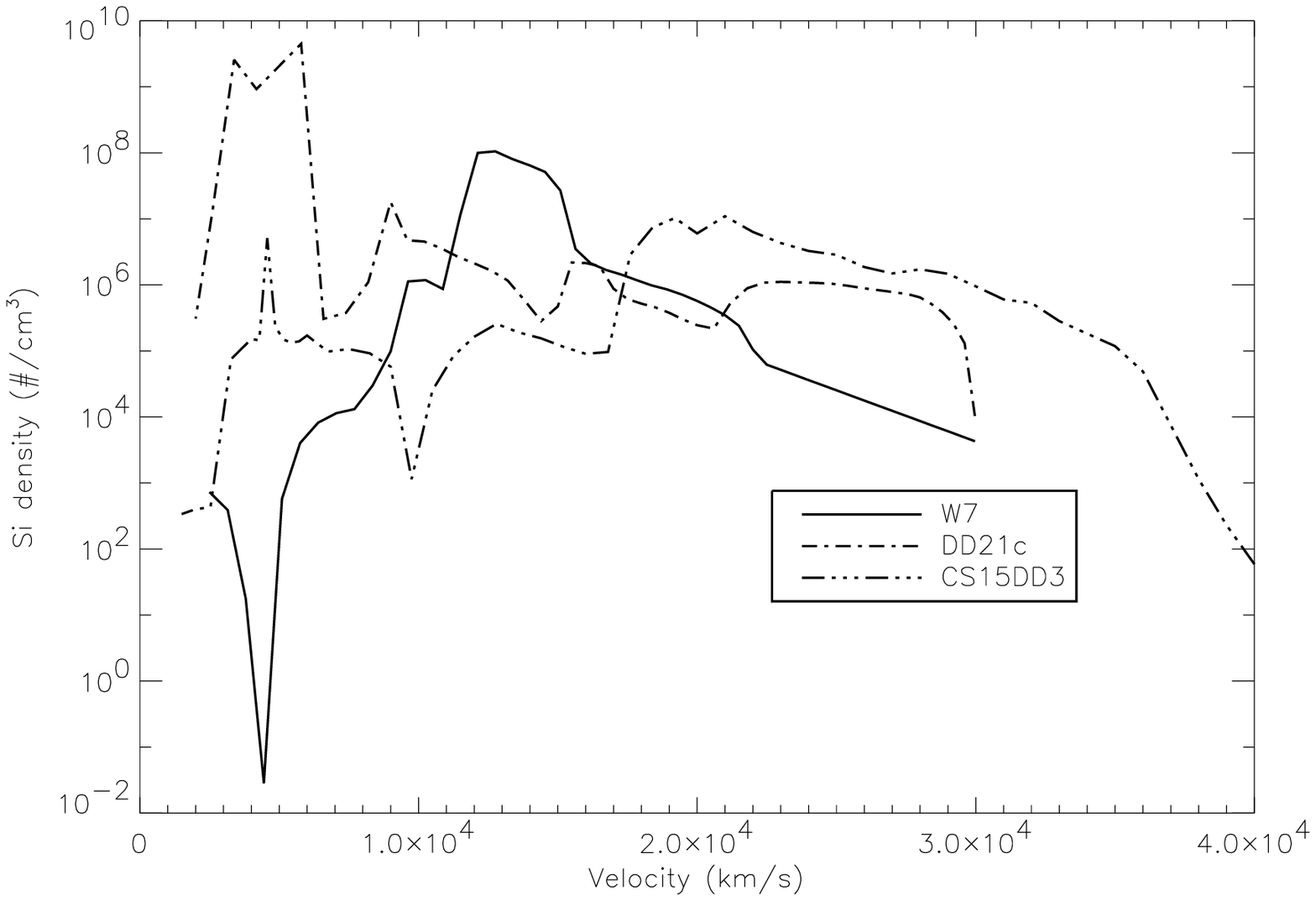}
\end{center}
\caption{Relative silicon number density for
\cdd~(triple dot-dashed line), W7~(solid line), and \dd~(dot-dashed
line) plotted against velocity.\label{fig:siden}} 
\end{figure}

\begin{figure} 
\begin{center} 
\leavevmode
\epsscale{0.95}
\plotone{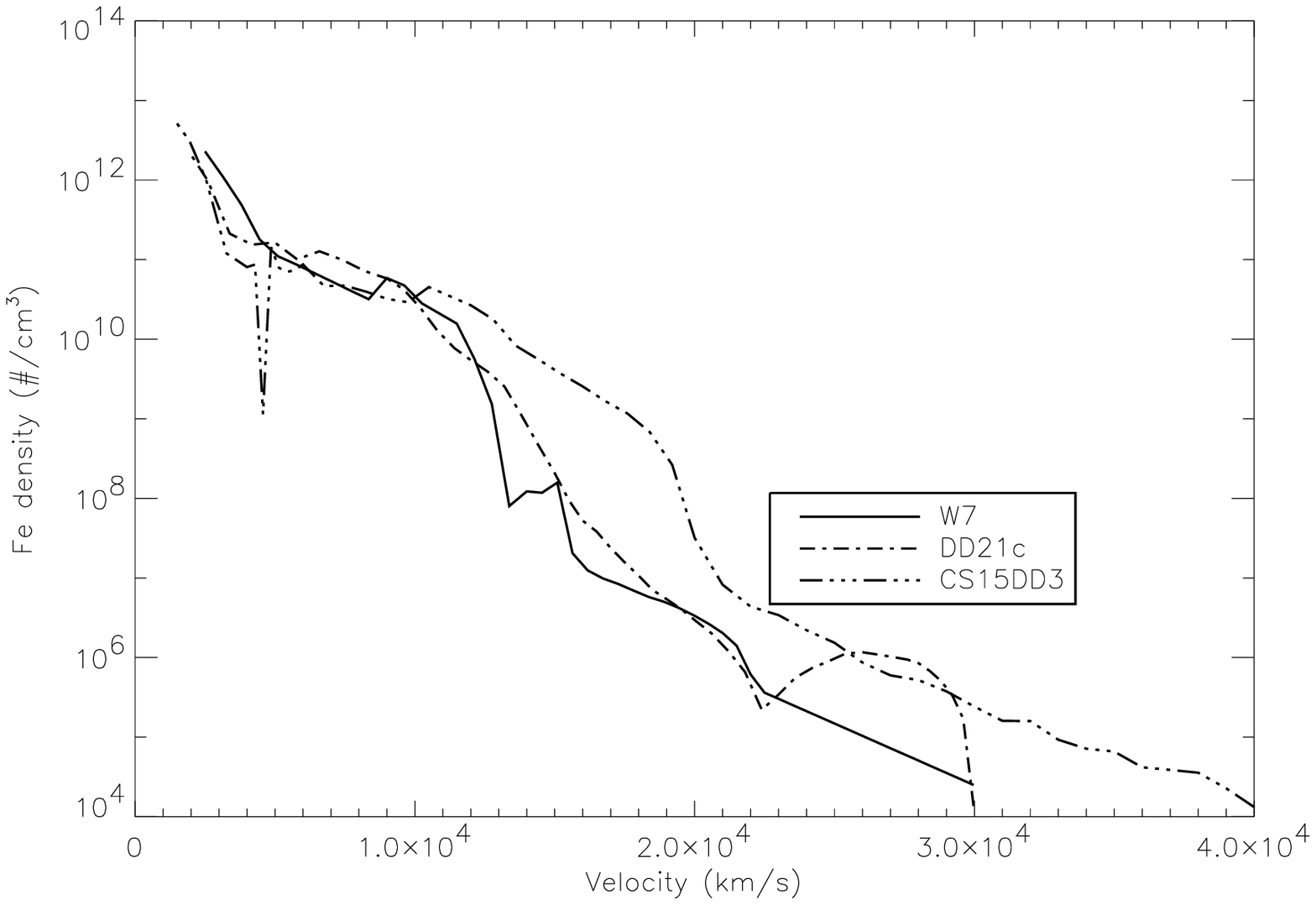}
\end{center}
\caption{Relative iron number density for
\cdd~(triple dot-dashed line), W7~(solid line), and \dd~(dot-dashed
line) plotted against velocity.\label{fig:feden}} 
\end{figure}

Figures~\ref{fig:siden} and \ref{fig:feden} compare the silicon and
iron number densities of the three models. Clearly, \dd\ and \cdd\ fit
the silicon feature well because they both have significant silicon
(nearly the same) up to $v=30,000$~\kmps. The smoothly rising iron
density of \cdd\ produces a better fit to the \ion{Fe}{2}
features than does the more complicated profile of \dd.

\section{Discussion}

We have calculated synthetic spectra from the delayed detonation models,
\dd\ and \cdd, to match the observed spectra of SN~1984A. The synthetic
spectra have successfully reproduced the high blue-shift absorption
features, and for \dd, the colors using the provided bolometric
luminosities. The extra column density in the intermediate mass elements
causes the lines to form with larger blue-shifts. We have previously
shown that line formation in this region of W7 is important in a study
of C+O metallicity \citep{lentzmet00}.
The variations in line velocities in \sneia\ with
otherwise normal spectra are most likely due to the addition of a small
amount of mass at high velocity, possibly containing freshly synthesized
material.  While the delayed detonation models we have used do show the
higher line blue-shifts and high-velocity densities, and the
deflagration model does not, we do not consider this as
\emph{conclusive} evidence that the variation is related to the
different   explosion mechanism. It may reflect a continuous variation
in the ejecta densities at high velocity within a single family of
models.  A delayed detonation model with less density at high
velocities, similar in outer density structure to W7, may fit the normal
velocity \sneia\ with the same or better quality than W7. In any case
it is clear that there must exist a family of ``normal'' SNe Ia and
that spectral modeling is the way to understand them.

\acknowledgments We thank K. Hatano and D. Casebeer for converting the
published observations to electronic format, and P. H\"oflich and
K. Nomoto for providing the explosion models in electronic format. We
thank the anonymous referee for suggestions which considerably improved the
presentation of this paper.  This work was supported in part by NSF
grant AST-9731450, NASA grant NAG5-3505, and an IBM SUR grant to the
University of Oklahoma; and by NSF grant AST-9720704, NASA ATP grant
NAG 5-8425, and LTSA grant NAG 5-3619 to the University of Georgia.
PHH was supported in part by the P\^ole Scientifique de Mod\'elisation
Num\'erique at ENS-Lyon. Some of the calculations presented in this
paper were performed at the San Diego Supercomputer Center (SDSC),
supported by the NSF, and at the National Energy Research
Supercomputer Center (NERSC), supported by the U.S. DOE.  We thank
both these institutions for a generous allocation of computer time.

\bibliography{refs,baron,sn1bc,sn1a,snii,gals,crossrefs}

\clearpage 

\end{document}